\documentclass[twocolumn]{aastex631}

\hypersetup{linkcolor=magenta,citecolor=cyan,filecolor=green,urlcolor=blue}

\usepackage{graphicx}
\usepackage{array}
\usepackage{booktabs}
\usepackage{float}
\usepackage{threeparttable}
\usepackage{tabularx}
\usepackage{amsmath}
\usepackage{enumitem}
\usepackage{multirow}
\usepackage[]{xcolor}
\usepackage{diagbox}
\usepackage[T1]{fontenc}

\begin{document}

\title{The nature of the accretion physics in quiescent black hole system LB-1}

\correspondingauthor{Tong Su, Erlin Qiao}
\email{sutong@bao.ac.cn, qiaoel@nao.cas.cn}

\author{Tong Su}
\affiliation{National Astronomical Observatories, Chinese Academy of Sciences, Beijing 100101, China}
\affiliation{School of Astronomy and Space Sciences, University of Chinese Academy of Sciences, 19A Yuquan Road, Beijing 100049, China}

\author{Erlin Qiao}
\affiliation{National Astronomical Observatories, Chinese Academy of Sciences, Beijing 100101, China}
\affiliation{School of Astronomy and Space Sciences, University of Chinese Academy of Sciences, 19A Yuquan Road, Beijing 100049, China}

\author{Song Wang}
\affiliation{National Astronomical Observatories, Chinese Academy of Sciences, Beijing 100101, China}
\affiliation{Institute for Frontiers in Astronomy and Astrophysics, Beijing Normal University, Beijing 102206, China}








\begin{abstract}
{LB-1 is a binary system that has drawn great attention since its discovery in 2019. The nature of the two components of LB-1 is not very clear, which however is suggested very possibly to be a  B-type star plus a black hole (BH). In this paper, we first calculate the wind mass-loss rate of the B-type star. We then calculate the mass capture rate by the BH, with which as the initial mass accretion rate, we calculate the truncation radius of the accretion disk and the corresponding emergent spectra of the accretion flow (comprising an inner advection-dominated accretion flow (ADAF) + an outer truncated accretion disk) within the framework of the disk evaporation model. It is found that the predicted truncation radius of the accretion disk with appropriate model parameters is consistent with observations inferred from the observed broad $\rm H_{\alpha}$ emission line. The predicted X-ray luminosity is definitely below the estimated upper limits with the sensitivity of \textit{Chandra} X-ray Observatory of the X-ray luminosity $\sim 2\times 10^{31}$ erg/s. Finally, we argue that if the disk evaporation model indeed reflects the intrinsic physics of the accretion flow, the value of the viscosity parameter $\alpha$ is constrained to be $\alpha \gtrsim 0.05$ (with BH mass being $68M_{\rm \odot}$), or $\alpha \gtrsim 0.003$ (with BH mass being $21M_{\rm \odot}$) to match the observed upper limit of the X-ray luminosity of LB-1.}
\end{abstract}



\section{Introduction} \label{sec:intro}

LB-1 (LS V+22 25) is discovered as a binary system by a radial-velocity monitoring campaign of Large Aperture Multi-Object Spectroscopic Telescope(LAMOST) in the Kepler K2-0 field of the 
sky \citep[]{2019Natur.575..618L}. The coordinate of LB-1 is $(l,b)=(188.23526^{\circ}, +02.05089^{\circ})$, where $l$ is Galactic longitude and $b$ is Galactic latitude. Besides the stellar absorption lines, 
a broad $\rm H_{\alpha}$ emission line is identified in LAMOST spectra, which is almost stationary. Subsequent observations by GTC/OSISRIS and Keck/HIRES confirmed the 
apparent periodic motion of the stellar absorption lines and the presence of the prominent broad $\rm H_{\alpha}$ emission line. Meanwhile, it is found that the $\rm H_{\alpha}$ emission line is moving in 
an anti-phase with much smaller amplitude compared with that of the stellar absorption lines. 
The orbital period of LB-1 is  $P=78.9\pm0.3 \ \mathrm{d}$. The semi-amplitude velocity of the stellar absorption line and the $\rm H_{\alpha}$ emission line are 
$K_{\mathrm{B}}=52.8\pm  0.7 \rm \ km\ s^{-1}$ and $K_{\mathrm{A}}=6.4\pm 0.8 \rm \ km \ s^{-1}$ respectively. 
Spectral modeling with TLUSTY indicates that LB-1 is a B-type main sequence star with a mass of $M_{\rm B}=8.2_{-1.2}^{+0.9}\ M_\odot$, an effective temperature 
$T_{\mathrm{eff}}=18100\pm820 \rm \ K$ and a surface gravity $\log{g}=3.43\pm0.15$. The dark companion of LB-1 is suggested to be a BH (with a minimum mass of $6.3_{-1.0}^{+0.4}M_\odot$) from the mass function measurement assuming an edge-on orbit. 
One of the very interesting features of LB-1 is the observed broad $\rm H_{\alpha}$ emission line with the full-width at half-maximum (FWHM) $\sim 240 \rm \ km \ s^{-1}$, which is too broad to be 
from an interloper M dwarf, surrounding nebulae or a circumbinary disk. It is suggested that the broad  $\rm H_{\alpha}$ emission line is from an accretion disk around the BH, 
which consequently can be used to trace the motion of the BH. 
According to the conservation of angular momentum, the mass of the BH can be calculated as
$M_{\mathrm{BH}}=(K_{\mathrm{B}}/K_{\mathrm{A}}) M_\mathrm{B}\approx68_{-13}^{+11}\ M_\odot$, which corresponds to an inclination angle of 
$15^{\rm o}$--$18^{\rm o}$ if the BH mass is measured with the method of mass function measurement. The derived small  inclination angle is consistent with the wine-bottle shape of the 
$\rm H_{\alpha}$ emission line \citep[]{2019Natur.575..618L}.

The estimated BH mass of $M_{\mathrm{BH}}\sim 68 M_\odot$ is questioned by some other groups \citep[e.g.][]{2020Natur.580E..11A,2020A&A...634L...7S,2020A&A...633L...5I,2020A&A...639L...6S}.
The debates on the BH mass are mainly from two aspects: (1) whether the $\rm H_{\alpha}$ emission line can trace the motion of the BH (or even not a BH). 
(2) the uncertainty for the mass measurements of the B-type star.
For example, based on the optical spectra observed by HERMES spectrograph \citep{Raskin2011}, 
\citet{2020Natur.580E..11A} argue that the observed radial-velocity measurements are resulted from the superposition of the stellar absorption 
line and a nearly static $\rm H_{\rm \alpha}$ emission line, so the  $\rm H_{\rm \alpha}$ emission line can not be used to trace the motion of the BH. In addition, it is argued that 
the mass of the B-type star is estimated to be $M_{\rm B}=4.2_{-0.7}^{+0.8}\ M_\odot$ by fitting the HERMES optical spectra with a local thermodynamic equilibrium (LTE) atmospheric 
model \citep[][]{El-Badry2020,Tkachenko2015}. The study in \citet[]{2020Natur.580E..11A} may alleviate the challenge for the presence of a very
massive BH at solar metallicity. Further, it is proposed that the companion is a fast rotating B-type star with a mass similar to that of the primary B-type star. Alternatively,  
the companion is still possible to be a BH, however the mass of the BH is suggested to be not greater than $\sim 50\ M_\odot$.
In \citet[][]{2020A&A...634L...7S}, the authors propose that the B-type star is a slightly evolved main sequence star with the mass $\sim$ 3--5 $M_\odot$ by jointly fitting the 
spectra of TNG/HARPS-N, GTC/HORuS and Keck/HIRES used in \citet[]{2019Natur.575..618L} with the non-LTE stellar atmosphere code \textsc{fastwind}. 
The mass of the BH is derived to be $\sim$ 4--5 $M_\odot$ by assuming an edge-on orbit with the method of mass function measurement.
In \citet[][]{2020A&A...633L...5I}, the authors reanalyze the Keck/HIRES spectra used in \citet[]{2019Natur.575..618L} with a different stellar atmosphere model
\citep[][]{2014A&A...565A..63I, 2018A&A...620A..48I}, and find that the B-type spectral star of LB-1 is a stripped helium star with a mass of $M_{\rm B}=1.1\pm 0.5 M_\odot$ rather than 
a B-type main sequence star. Further, the authors suggest that the dark companion of LB-1 is a compact object with a minimum mass of 2--3 $M_\odot$ based on the mass function measurement,
which could be a mass gap BH, a massive neutron star, or even a main sequence star.
Combing the observed spectral data of HERMES and FEROS spectrographs, \citet[][]{2020A&A...639L...6S} also suggests that the primary of LB-1 is a stripped helium star with a mass 
of $\sim 1\ M_\odot$. Meanwhile, it is suggested that the dark companion is a $\rm Be$ star, which contributes about $45\%$ optical flux.
Assuming a typical value of $7 \pm 2\ M_\odot$ for the $\rm Be$ star (i.e., the dark companion), the mass of the primary of LB-1 is 
further constrained to be a mass of $1.5\pm 0.4\ M_\odot$, and the orbital inclination angle of LB-1 is estimated to be $39^{\rm o} \pm 4^{\rm o}$.

In order to respond to the debates on the mass of the dark companion, \citet[][]{2020ApJ...900...42L} conduct a new study, analyzing the $\rm Pa\beta$
and $\rm Pa\gamma$ emission lines observed with Calar Alto high- Resolution search for M dwarfs with Exo-earths with Near-infrared and optical Echelle Spectrographs (CARMENES) mounted on 
the 3.5 m telescope at the Calar Alto Observatory. 
The phase-averaged $\rm Pa\beta$ and $\rm Pa\gamma$ emission lines show a cleaner, double-peaked disk-like profile, which is clearer 
for exploring the properties of the dark companion compared with the $\rm H_{\alpha}$ emission line
{due to the complex distortion of $\rm H_{\alpha}$ emission line.}
The authors measured the shift of the line center of the $\rm Pa\beta$ and $\rm Pa\gamma$ emission lines, showing a perfect anti-phase motion with the stellar absorption lines. 
The semi-amplitude velocity of the $\rm Pa\beta$ and $\rm Pa\gamma$ emission line is in the range of 8--13 $\rm \ km\ s^{-1}$. Further, based on the line profile, it is proved that the $\rm Pa\beta$ 
and $\rm Pa\gamma$ emission lines can trace the dark companion, ruling out the circumbinary disk and the hierarchical triple cases.
Combing the semi-amplitude of the stellar absorption line for the primary B-type star, and the $\rm Pa\beta$ and $\rm Pa\gamma$ emission line for the dark companion, the inferred 
mass ratio of the dark companion to the primary B-type star is $5.1\pm 0.8$.
Finally, several possibilities for the nature of the two components of LB-1, in particular, a B-type main sequence star plus a BH (B+BH), a stripped helium star plus a $\rm B{\rm e}$ 
star ($\rm B_{\rm He}$+$\rm B{\rm e}$) are discussed \citep[][]{2020ApJ...900...42L}.

Recently,  \citet{Lennon2021} test the {scenarios} of B+BH and $\rm B_{\rm He}$+$\rm B{\rm e}$ by fitting the UV-optical spectra obtained from 
Space Telescope Imaging Spectrograph (STIS) and the IR spectrum spectra obtained from Wide Field Camera 3 (WFC3) onboard Hubble Space Telescope (HST).
It shows that the B+BH model is more preferred, in particular for the explanation for the UV spectrum. In the B+BH scenario, the mass of the B-type star is fitted to be 
$5.1_{-1.4}^{+1.8}\ M_\odot$, and the mass of the BH is derived to be $21_{-8}^{+9}\ M_\odot$ by assuming a mass ratio of the dark companion to the primary $5.1\pm0.1$ as
suggested in \citet[][]{2020ApJ...900...42L}.

In this paper, we investigate the properties of the accretion flow with the scenario of B+BH for LB-1. Specifically, we calculate the wind mass-loss rate of the B-type star, and 
the mass rate captured by the BH. Using this as the initial mass accretion rate,
we calculate the truncation radius of the accretion disk and the corresponding
emergent spectra of the accretion flow (comprising an inner advection-dominated accretion flow (ADAF) + an outer truncated accretion disk) within the framework of the disk evaporation model.
The generally predicted features by the disk evaporation model of the accretion flow, such as the truncation radius of the accretion disk, the very dim X-ray emission (non-detection), as well as 
the nearly neglected UV-optical emission compared to that of the B-type star, are consistent with observations of LB-1. Finally, we discuss the effect of viscosity parameter $\alpha$ on the 
multi-band emission of the accretion flow. 
In Section. \ref{sec:model}, we calculate the wind mass-loss rate of the B-type star and the mass rate captured by the BH in LB-1. In Section. \ref{sec:result}, we briefly introduce the disk evaporation model
and the corresponding predicted features as applied in LB-1. The discussions are in Section. \ref{sec: discuss}, and the conclusions are in Section. \ref{sec: conclude}.


{\section{The models} \label{sec:model}
\subsection{Wind mass-loss rate of the B-type star and the mass rate captured by the BH} \label{sec: model_rate}
We first calculate the wind mass-loss rate of the B-type star, following \citep{2000A&A...362..295V},
\begin{equation}\label{sec2:lossrate}
\begin{aligned}
\log \dot{M}_\mathrm{win}=&-6.688+2.21\log \left(L_\mathrm{B} / 10^5\right)-1.339\log \left(M_\mathrm{B} / 30\right) \\
&-1.601\log \left(\frac{v_{\infty} / v_{\text {esc}}}{2.0}\right)+1.07 \log \left(T_{\text {eff }} / 20000\right),
\end{aligned}
\end{equation}
where  $\dot{M}_\mathrm{win}$ is the wind mass-loss rate in units of $M_\odot \ \mathrm{yr^{-1}}$, $L_\mathrm{B}$ and $M_\mathrm{B}$ are the bolometric luminosity and the mass 
of the B-type star respectively in solar units, $T_\mathrm{eff}$ is the effective temperature at the surface of the B-type star, and $v_\infty/v_\mathrm{esc}$ is the ratio of the terminal velocity  
$v_\infty$ to the effective escape velocity at the stellar surface $v_\mathrm{esc}$ of the wind, $v_\mathrm{esc}=\sqrt{\frac{2GM_\mathrm{B}(1-\Gamma_\mathrm{e})}{R_\mathrm{B}}}$, where $G$ is the gravitational constant, $R_\mathrm{B}$ is the radius of the B-type star, and 
$\Gamma_e \equiv L_{\rm B}/L_\mathrm{Edd} = \kappa_\mathrm{e} L_\mathrm{B}/(4\pi c G M_\mathrm{B})$ is the Eddington parameter. Equation (\ref{sec2:lossrate}) is valid for $T_{\rm eff}$ in the range of
$12500\ \mathrm{K}<T_\mathrm{eff}<22500\ \mathrm{K}$; in this temperature range $v_\infty/v_\mathrm{esc}=1.3$.
The wind velocity near the BH $v_\mathrm{win}$ can be expressed as, 
\begin{equation}\label{sec2:vwind}
v_\mathrm{win}(A)=v_{\infty}\left(1-\frac{R_\mathrm{B}}{A}\right)^\beta,
\end{equation}
where $A$ is the separation of the binary system, calculated as $A=\left[\frac{G\left(M_{\mathrm{B}}+M_{\mathrm{BH}}\right) P^2}{4 \pi^2}\right]^{1/3}$ with the Kepler's third law.
$M_{\rm BH}$ is the mass of the BH, and $P$ is the orbital period. The index $\beta$ is $\sim 1$ for OB supergiants \citep{2000A&A...362..295V}. 

The mass capture radius can be estimated in terms of the Bondi-Hoyle prescription,
\begin{equation}\label{sec2:Rcap}
R_{\mathrm{cap}}=\frac{2 G M_\mathrm{BH}}{v_\mathrm{win}^2},
\end{equation}
and the mass rate captured by the BH, $\dot{M}_\mathrm{cap}$, can then be expressed as, 
\begin{equation}\label{sec2:Mdotcap}
\dot{M}_{\text {cap }}=\frac{\pi R_{\text {cap }}^2}{4 \pi A^2} \dot{M}_{\text {win}}.
\end{equation} 
Combing Equations (\ref{sec2:lossrate}), (\ref{sec2:vwind}), (\ref{sec2:Rcap}) and  (\ref{sec2:Mdotcap}), we calculate  $\dot{M}_\mathrm{cap}$ by specifying $L_{\rm B}$, $M_{\rm B}$, $R_{\rm B}$,
$T_{\rm eff}$, $M_{\rm BH}$ and $P$ measured in LB-1.  
In this paper, two groups of system parameters of LB-1, named G1 and G2, are used for calculating $\dot{M}_\mathrm{cap}$.  
One can refer to Table \ref{tab:para} for the data of G1 and G2 for details. 
The data of G1 is from \citet[][]{2019Natur.575..618L}, which is the original paper for LB-1, and the data of G2 is from \citet[][]{Lennon2021} (note that the period $P$ and the semi-amplitude $K_\mathrm{B}$ are taken from \citet{2020ApJ...900...42L}), in which by 
fitting the UV-optical spectra obtained from STIS and the IR spectrum spectra obtained from WFC3 onboard HST, the authors 
find the fitting results of B+BH model is more preferred than that of the $\rm B_{\rm He}$+$\rm B{\rm e}$ model for LB-1.
We list the value of $\dot{M}_\mathrm{cap}$ in column (6) of Table \ref{tab:para22}, in which $\dot M_{\rm cap}$ is $3.08\times 10^{-11}\ M_\odot/$yr for G1
and $4.32\times 10^{-13}\ M_\odot/$yr for G2. 
The mass rate captured by the BH $\dot{M}_\mathrm{cap}$, can be regarded as the initial mass accretion rate fed for the BH, denoted as $\dot M$ hereafter.
In this paper, we define a dimensionless mass accretion rate, i.e., $\dot m\equiv \dot M/\dot M_{\rm Edd}$  (with $\dot M_{\rm Edd}=L_{\rm Edd}/{0.1c^2}$=$1.39 \times 10^{18} M_{\rm BH}/M_\odot \rm \ g \ s^{-1}$) 
for convenience. We list $\dot m$ in column (7) of Table \ref{tab:para22}. The value of $\dot m$ is $2.05\times10^{-5}$ for G1 and $9.32\times10^{-7}$ for G2.
We also list the value of the separation of the binary system $A$, the capture radius $R_{\rm cap}$, the wind mass-loss rate of the B-type star $\dot M_{\rm win}$, 
and  ${\rm sin}\ i$ of the binary system (with $i$ being the inclination angle of the orbit of the binary system), which can be found in Table \ref{tab:para22} for details.

\begin{deluxetable*}{cccccccccc}
\tabletypesize{\scriptsize}
\tablewidth{0pt} 
\tablecaption{ System parameters of LB-1, taken from \citet[]{2019Natur.575..618L}(G1) and \citet[]{Lennon2021}(G2) respectively. $L_{\rm B}$, $M_\mathrm{B}$, $R_\mathrm{B}$, $T_\mathrm{eff}$, $\log{g}$, and $K_\mathrm{B}$ are the luminosity, mass, radius, effective temperature, logarithm of the surface gravity, and the semi-amplitude velocity of the B-type star, and the $M_\mathrm{BH}$ is the BH mass, all scaled to solar units.\label{tab:para} }
\tablehead{
\colhead{System parameter} & 
\colhead{$L_{\rm B}$ ($L_\odot$)} & 
\colhead{$M_\mathrm{B}$ ($M_\odot$)}  & 
\colhead{$R_\mathrm{B}$ ($R_\odot$)} & 
\colhead{$T_\mathrm{eff}$ (K)} &
\colhead{$\log{g}$} & 
\colhead{$K_\mathrm{B}$ (km/s)} & 
\colhead{$P$ (days)} & 
\colhead{$\sin{i}$} &
\colhead{$M_\mathrm{BH}$ ($M_\odot$)}  \\
\colhead{ } & \colhead{(1)} & \colhead{(2)} & \colhead{(3)} & \colhead{(4)} & \colhead{(5)} & \colhead{(6)} & \colhead{(7)} & \colhead{(8)} & \colhead{(9)} 
} 
\startdata 
{G1} & 7000 & 8.2 & 9.0 & 18000 & 3.43 & 52.8 & 78.9 & 0.28 & 68 \\
{G2} & 1698 & 5.2 & 6.0 & 15300 & 3.6  & 52.6 & 78.9 & 0.43 & 21 \\
\enddata
\end{deluxetable*}

\begin{deluxetable*}{ccccccccc}
\tabletypesize{\scriptsize}
\tablewidth{0pt} 
\tablecaption{Derived parameters of LB-1. Columns (1) and (2) are the separation of the binary system, $A$, in units of $\rm cm$ and $R_\mathrm{S}$ 
$R_{\rm S}$ respectively. Columns (3) and (4) are the capture radius, $R_{\rm cap}$,
in units of $\rm cm$ and $R_{\rm S}$ respectively. Column (5) is the wind mass-loss rate of the B-type star, $\dot M_{\rm win}$, in units of $M_\odot/$yr, and  Column (6) is 
the mass rate captured by the BH,  $\dot M_{\rm cap}$, in units of $M_\odot/$yr.  Column (7) is the mass accretion rate fed to the BH in units of $\dot M_{\rm Edd}$.
Column (8) is ${\rm sin}i$ of the the binary system with $i$ being the inclination angle of the orbit.
\label{tab:para22}}
\tablehead{
\colhead{Derived parameter} & \colhead{$A$(cm)}& \colhead{$A(R_\mathrm{S})$}& \colhead{$R_\mathrm{cap}$(cm)} & \colhead{$R_\mathrm{cap}(R_\mathrm{S})$} & \colhead{$\dot{M}_\mathrm{win}(M_\odot / \mathrm{yr})$} & \colhead{$\dot{M}_\mathrm{cap}(M_\odot / \mathrm{yr})$} &\colhead{$\dot{m}(\dot m\equiv \dot{M}/\dot{M}_\mathrm{Edd})$} & \colhead{$\sin{i}$} \\
\colhead{ } & \colhead{(1)} & \colhead{(2)} & \colhead{(3)} & \colhead{(4)} & \colhead{(5)} & \colhead{(6)} & \colhead{(7)} & \colhead{(8)}
} 
\startdata 
$\mathrm{G1}$ & $2.28\times 10^{13}$ & $1.14\times10^6$ & $3.32\times 10^{12}$ & $1.66\times10^5$ & $5.81\times 10^{-9}$ & $3.08\times 10^{-11}$ & $2.05\times10^{-5}$ & 0.28 \\
$\mathrm{G2}$ & $1.60\times 10^{13}$ & $2.58\times10^6$ & $1.06\times 10^{12}$ & $1.71\times10^5$ & $3.93\times 10^{-10}$ & $4.32\times 10^{-13}$ & $9.32\times10^{-7}$ & 0.43 \\
\enddata
\end{deluxetable*}

{\subsection{A summary of the disk evaporation model} \label{sec: model_spectra}

With the mass accretion rate $\dot m$ calculated in Section \ref{sec: model_rate}, we further investigate the geometry of the accretion flow and the corresponding emergent spectra. We assume that the 
accretion flow is in the form of the standard accretion disk initially \citep[][]{Shakura1973}, since strong evidence indicates the presence of an accretion disk around the 
BH as inferred from modeling the profiles of $\rm H_{\alpha}$ \citep[][]{2019Natur.575..618L}, $\rm Pa\beta$, and $\rm Pa\gamma$ emission lines \citep[][]{2020ApJ...900...42L}.
In general, there exists a critical mass accretion rate $\dot m_{\rm crit}$ (accretion rate in units of $\dot M_{\rm Edd}$). If $\dot m$ is greater than $\dot m_{\rm crit}$, the standard 
accretion disk will extend down to the innermost stable circular orbit (ISCO) of the BH. While, if $\dot m$ is less than $\dot m_{\rm crit}$, the accretion disk will truncate at some radius.
The truncation radius generally decreases with increasing mass accretion rate.
Several models have been proposed for $\dot m_{\rm crit}$ and the truncation radius of the accretion disk.
One of the most promising models for $\dot m_{\rm crit}$ and the truncation of the accretion disk} is 
the disk evaporation model \citep[][]{Meyer1994,Liubf1999,Meyer2000a,Meyer2000b,Liubf2002,Qiao2009,Qiao2010,Taam2012}. In \citet[][]{Taam2012}, the authors summarized the 
main results of the disk evaporation model, giving general formulae for  $\dot m_{\rm crit}$ and the truncation radius of the accretion disk. We list $\dot m_{\rm crit}$ and the truncation 
radius $r_{\rm tr}$ as follows,
\begin{equation}
\dot m_{\rm crit} \approx 0.38\alpha^{2.34}\beta^{-0.41},
\end{equation}
\begin{equation}\label{equ:rtr}
r_{\mathrm{tr}} \approx 17.3 \dot{m}^{-0.886} \alpha^{0.07} \beta^{4.61} ,
\end{equation}
where $\alpha$ is the viscosity parameter, $\beta$ is the magnetic parameter (with magnetic pressure $p_{\rm m}={B^2/{8\pi}}=(1-\beta)p_{\rm tot}$, $p_{\rm tot}=p_{\rm gas}+p_{\rm m}$).
$r_{\rm tr}$ is the truncation radius in units of Schwarzschild radius $R_{\rm S}$
(with $R_{\rm S}=2GM_{\rm BH}/c^2\approx 2.95\times 10^{5}\ M_{\rm BH}/M_\odot\ \rm cm$) \footnote {Note
that the derived formulae for $\dot m_{\rm crit}$ and $r_{\rm tr}$ are independent on the black hole mass, so can be applied to both stellar-mass BH and supermassive BH.}.

}

\begin{figure}[h!]
\centering
\includegraphics[width=0.5\textwidth]{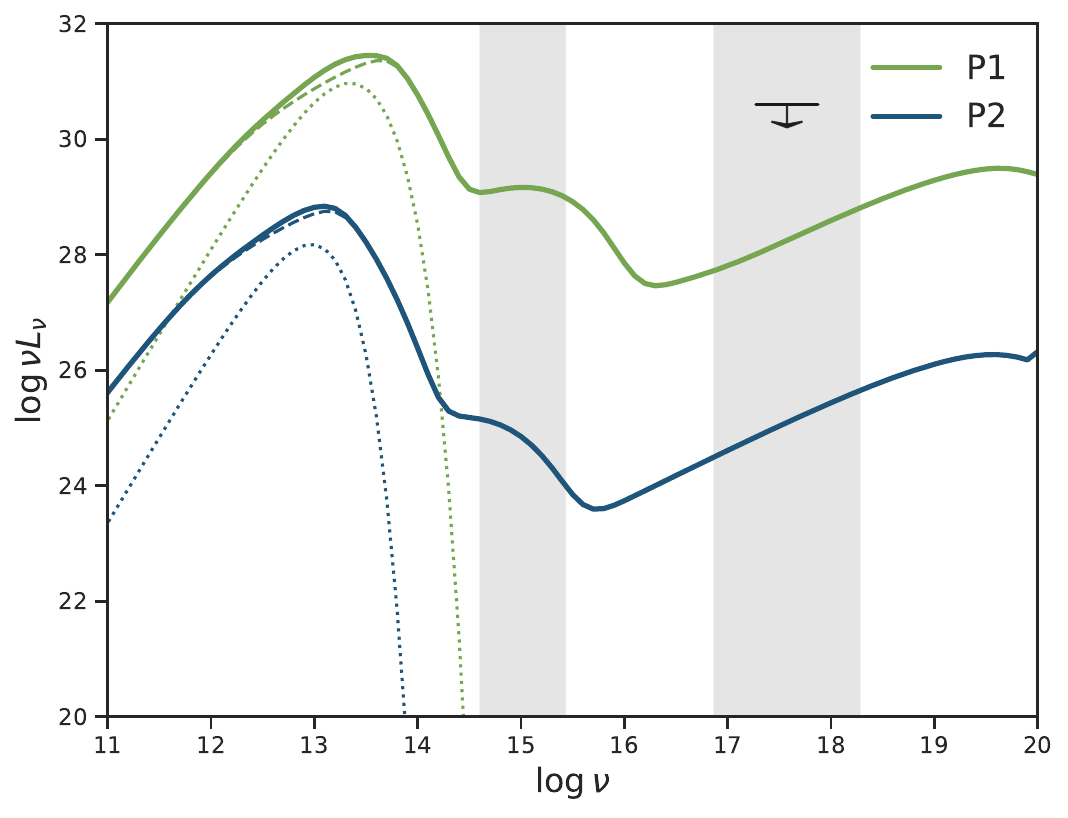}
\caption
{Emergent spectra of the two groups of model parameters (P1: $[m, \dot{m}, \alpha, \beta] = [68, 2.05\times10^{-5}, 0.3, 0.75]$; P2: $[m, \dot{m}, \alpha, \beta] = [21, 9.32\times10^{-7}, 0.3, 0.50]$.). 
The green solid line is the total emergent spectra of model parameter P1, and the blue solid line is the total emergent spectra of model parameter P2. The green (blue) dashed line and green (blue) dotted line present the emission of the inner ADAF and the truncated accretion disk for model parameters P1 and P2 respectively.
The two shaded areas, from left to right, represent the UV-optical band (1100\AA--7600\AA) and the X-ray band (0.3--8 keV) respectively. The black downward arrow represents the upper limit of the X-ray luminosity of LB-1 given by \textit{Chandra} X-ray observatory.
\label{fig:SED1}}
\end{figure}

{\section{result} \label{sec:result}
\subsection{The truncation radius of the accretion disk}\label{sec:truncation}
According to Equation (\ref{equ:rtr}), the truncation radius of the accretion disk $r_{\rm tr}$ can be calculated by specifying $\dot m$, $\alpha$ and $\beta$.
It can be seen from Equation (\ref{equ:rtr}) that $r_{\rm tr}$ is very weakly dependent on $\alpha$. Initially, we fix $\alpha=0.3$ as adopted in several literatures
for the quiescent BH X-ray binaries \citep[e.g.][]{Yuan2005,Zhanghui2010}. Meanwhile, the effect of $\beta$ on $r_{\rm tr}$ is very strong as has been discussed in \citet{Meyer2002}, \citet{Qian2007}, and \citet{Taam2012}. We test the effect of $\beta$
on $r_{\rm tr}$ for a given value of $\dot m$ derived from the system parameters G1 and G2 of LB-1 respectively. 

For the system parameter G1, $\dot m$ is calculated to be $2.05\times10^{-5}$. Substituting $\dot m$ into Equation (\ref{equ:rtr}), $r_{\rm tr}$ are calculated to be $8.10\times10^4$, $6.01\times10^4$ and $4.37\times10^4$ $R_\mathrm{S}$ for $\beta=$0.8, 0.75, and 0.70 respectively. Assuming the truncated accretion disk is Keplerian in the angular direction, at $r_{\rm tr}$ the projected velocity
in the line of sight can be calculated as $v_{\rm K, tr}{\rm sin}i$, which are $\sim$ 209.57, 243.18, 285.10 km/s for $\beta=$0.8, 0.75, and 0.70 respectively. We further calculate
the effective temperature at the surface of the accretion disk $T_{\rm eff}(r_{\rm tr})$ \citep[][]{Shakura1973,Frank2002}, which are $\sim$ 303.72, 379.55, and 481.66 K respectively.
Hydrogen in this temperature range is neutral, which ensures the production of hydrogen emission line.
As for the system parameter G2, $\dot m$ is calculated to be $9.32\times10^{-7}$. In this case, $r_{\rm tr}$ are calculated to be $2.23\times10^5$, $1.43\times10^5$ and $8.82\times10^4$ $R_\mathrm{S}$ for 
$\beta=$0.55, 0.5, and 0.45 respectively. Likewise, assuming the angular velocity of the truncated accretion disk is the Keplerian velocity, at $r_{\rm tr}$, 
$v_{\rm K, tr}{\rm sin}\ i$ are $\sim$ 194.56, 242.37, 308.99 km/s for $\beta$=0.55, 0.5, and 0.45 respectively, and the corresponding  
$T_{\rm eff}(r_{\rm tr})$  are $\sim$ 88.17, 122.56 and 176.37 K respectively. For clarity, one can refer to Table \ref{table:para3} for the information related to the 
truncated radius of the accretion disk.

Clearly, with $\dot m=2.05\times10^{-5}$ (based {on} G1 data) and $\beta=0.75$, or $\dot m=9.32\times10^{-7}$ (based on G2 data) and $\beta=0.5$,
the predicted projected velocity of the truncated disk in the line of sight matches very well with the observed FWHM of the $\rm H_{\alpha}$ emission line of $\sim 240\ \rm km/s$. 
To more easily compare with observations, in Section \ref{sec:sp}, we calculate the corresponding emergent spectra of the accretion flow with $m=68$, $\dot m=2.05\times10^{-5}$, $\alpha=0.3$ and $\beta=0.75$, as well as 
$m=21$, $\dot m=9.32\times10^{-7}$, $\alpha=0.3$ and $\beta=0.5$ respectively.

\begin{deluxetable*}{cccccccc}
\tabletypesize{\scriptsize}
\tablewidth{0pt} 
\renewcommand{\arraystretch}{1.2}
\tablecaption{Parameters of the accretion flow. Column (1) is the mass accretion rate $\dot m$, Column (2) is the viscosity parameter $\alpha$,  which is fixed to be  $\alpha=0.3$. 
Column (3) is the magnetic parameter $\beta$. Column (4) is the calculated truncation radius of the accretion disk from Equation (\ref{equ:rtr}). Column (5) is the corresponding 
projected velocity in the line of sight at the truncation radius. Column (6) is the corresponding effective temperature at the surface of the accretion disk at the truncation radius.
\label{table:para3}}
\tablehead{
Parameters of the accretion flow  & $\dot{m}$ & $\alpha$ & $\beta$ & $r_\mathrm{tr}$ & $v_\mathrm{K,tr}\sin{i}$ & $T_{\rm eff}(r_\mathrm{tr})$ & $R_\mathrm{tr}<R_\mathrm{cap}$? \\
\multirow{2}{*}{\diagbox[width=4.4cm, height=0.95cm]{}{}} & &  &  & $(R_\mathrm{S})$ & 
($\mathrm{km/s}$) & (K) &  \\
 & (1) & (2) & (3) & (4) & (5)& (6)& (7)
}
\startdata 
&&& 0.80& $8.10\times10^4$ & 209.57 & 303.72 & True \\
&$2.04\times10^{-5} $~(G1)&0.3& 0.75& $6.01\times10^4$ & 243.18 & 379.55 & True \\
&&& 0.70& $4.37\times10^4$ & 285.10 & 481.66 & True \\
\hline
&&& 0.55& $2.23\times10^5$ & 194.56 & 88.17 & False \\
&$9.07\times10^{-7}$~ ({G2})&0.3& 0.50& $1.43\times10^5$ & 242.37 & 122.56 & True \\
&&& 0.45& $8.82\times10^4$ & 308.99 & 176.37 & True \\
\enddata
\end{deluxetable*}

\subsection{The emergent spectra}\label{sec:sp}

Under the framework of the disk evaporation model, 
inside the truncation radius of the accretion disk, the accretion flow will be in the form of advection-dominated accretion flow (ADAF).
In this case, the geometry of the accretion flow will be an inner ADAF + an outer truncated accretion disk. In this paper, we calculate the emergent spectrum of the accretion flow by combing the emission from the inner ADAF and the outer truncated accretion disk. As for the ADAF, we adopt the self-similar solution for the structure \citep[][]{Narayan1994,1995ApJ...444..231N, 1995ApJ...452..710N, 1995Natur.374..623N},
and the multi-scattering method of the seed photons (including bremsstrahlung, synchrotron radiation of ADAF itself) in the 
optically thin, hot gas \citep{1990MNRAS.245..453C,1997ApJ...489..791M, Qiao2010,2013ApJ...764....2Q}. 
One can refer to \citet[][]{Qiao2010,2013ApJ...764....2Q} or \citet[][]{1997ApJ...489..791M} for the calculation of the emergent spectra of ADAF for details.

In Figure \ref{fig:SED1}, we plot the emergent spectra of the accretion flow with 
$m=68$, $\dot m=2.05\times10^{-5}$, $\alpha=0.3$ and $\beta=0.75$ (hereafter model parameter P1), and $m=21$, $\dot m=9.32\times10^{-7}$,  $\alpha=0.3$ and $\beta=0.5$ (hereafter model parameter P2), 
Specifically, in Figure \ref{fig:SED1}, the green solid line is the total emergent spectrum of P1. The green dashed line 
is the emission from the inner ADAF, and the green dotted line is the emission from the outer truncated accretion disk with truncation radius $R_{\rm tr}=6.01\times 10^4$ $R_\mathrm{S}$ calculated from
Equation (\ref{equ:rtr}). 
{On the other hand,} the blue solid line is for the total emergent spectrum of P2. The blue dashed line 
is the emission from the inner ADAF, and the blue dotted line is the emission from the outer truncated accretion disk with truncation radius $R_{\rm tr}=1.46\times 10^5$ $R_\mathrm{S}$ calculated from
Equation (\ref{equ:rtr}). 
The two shaded areas denote the UV-optical band (1100\AA-7600\AA) and the X-ray band (0.3-8 keV) respectively, and
the black downward arrow represents the upper limit of X-ray luminosity in the band of 0.3--8 keV given by the \textit{Chandra} X-ray observations.

It can be seen that for model parameter P1, the predicted X-ray luminosity is roughly two orders of magnitude lower than the estimated upper limit of X-ray luminosity, i.e., 
$L_{\rm X, BH}\sim 2\times 10^{31}$ erg/s, which is consistent with observations. For model parameter P2, the case is similar, the predicted X-ray luminosity is roughly four orders of magnitude lower than the upper limit of the X-ray luminosity, also well consistent with observations.
The emission of the truncated accretion disk is mainly in the infrared band, which is comparable with the emission of ADAF. However, the emission from both the truncated accretion disk and the ADAF in infrared are not well constrained by current observations.

As we can see from Equation (\ref{equ:rtr}), the truncation radius of the accretion disk $r_{\rm tr}$ is very weakly dependent on the value of $\alpha$. 
However, the structure and the bolometric luminosity of ADAF are very sensitive to $\alpha$. In general, the bolometric luminosity of ADAF is proportional to 
$\alpha^{-2}$ \citep[e.g.][]{1995ApJ...444..231N,Mahadevan1997,Qiao2009,Lijiaqi2023MNRAS}.
We test the effect of $\alpha$ on the predicted X-ray luminosity. 
In the left panel of Figure \ref{fig:alphas1}, based on the model parameter P1, fixing $m=68, \dot m=2.05\times10^{-5}$ and $\beta=0.75$, we calculate the emergent spectra
for different $\alpha$. From the bottom up, the pink, yellow, green, and blue solid lines are the emergent spectra for $\alpha=0.3, 0.1, 0.05, 0.02$ respectively. 
The dotted lines are the emission from the truncated accretion disk with a nearly unchanged truncation radius, i.e., 
$R_{\rm tr}\approx 6.0-5.0\times 10^4$ {$R_\mathrm{S}$} for $\alpha=0.3-0.02$. And the dashed lines are the emission from the inner ADAF.
It can be seen that the spectra shift upward systemically with decreasing $\alpha$. For $\alpha=0.02$, the X-ray luminosity between 0.3--8 keV $L_{\rm X,BH}$
is $\sim 1.5\times 10^{32}$ erg/s, exceeding the estimated upper limit of X-ray luminosity of $L_{\rm X}\sim 2\times 10^{31}$ erg/s given by the \textit{Chandra} X-ray observatory. while for $\alpha=0.05$, the X-ray luminosity $L_{\rm X,BH}\sim 6\times 10^{30}$ erg/s is well below the estimated upper limits of the X-ray luminosity. 
Therefore, $\alpha$ can be roughly constrained to be greater than $\sim 0.05$. 
As a comparison, we plot the emergent spectra of the corresponding B-type star (assuming a single black body) based on the system parameter G1, in which the spectrum is calculated as 
$L_{\rm \nu}=4\pi R_{\rm B}^2 \pi B_{\nu}(T_{\rm eff})$ with $R_{\rm B}$ being the radius and $T_{\rm eff}$ being the effective temperature of the B-type star, denoted as the black solid line.
It is clear that the UV-optical band emission is completely dominated by the  
B-type star, and the emission from the accretion flow in the UV-optical band can be completely neglected, which is consistent with the spectral fitting method adopted for LB-1 in \citet[]{2019Natur.575..618L}.
In the right panel of Figure \ref{fig:alphas1}, based on the model parameter P2, fixing  $m=21$, $\dot m=9.32\times10^{-7}$ and $\beta=0.5$, we calculate the emergent spectra for $\alpha=0.3, 0.01, 0.003, 0.001$, following the same line convention. It can be seen that the spectra shift upward systemically with decreasing $\alpha$. 
Specifically, for $\alpha=0.003$, $L_{\rm X,BH}$ is $\sim 2.4\times 10^{30}$ erg/s,
which is well below the estimated upper limits of the X-ray luminosity.
While for $\alpha=0.001$, $L_{\rm X,BH}$ is $\sim 2.2\times 10^{32}$ erg/s, 
exceeding the estimated upper limits of the X-ray luminosity.
Therefore, the value of $\alpha$ can be roughly constrained to be greater than $\sim 0.003$.
The black solid line denotes the spectra of the B-type star, based on the system parameter G2 for $R_{\rm B}$ and $T_{\rm eff}$. Similarly, the UV-optical band emission is also completely dominated by the B-type star. This is consistent 
with the fitting in \citet{Lennon2021} for the UV-optical spectra, in which the emission from the accretion flow is neglected. One can refer to Table \ref{tab:para3} for details on $L_\mathrm{BH, X}$ and $L_{\rm UV-Opt, B}/L_{\rm UV-Opt, BH}$ for different $\alpha$.}

\begin{figure*}
\includegraphics[width=2\columnwidth]{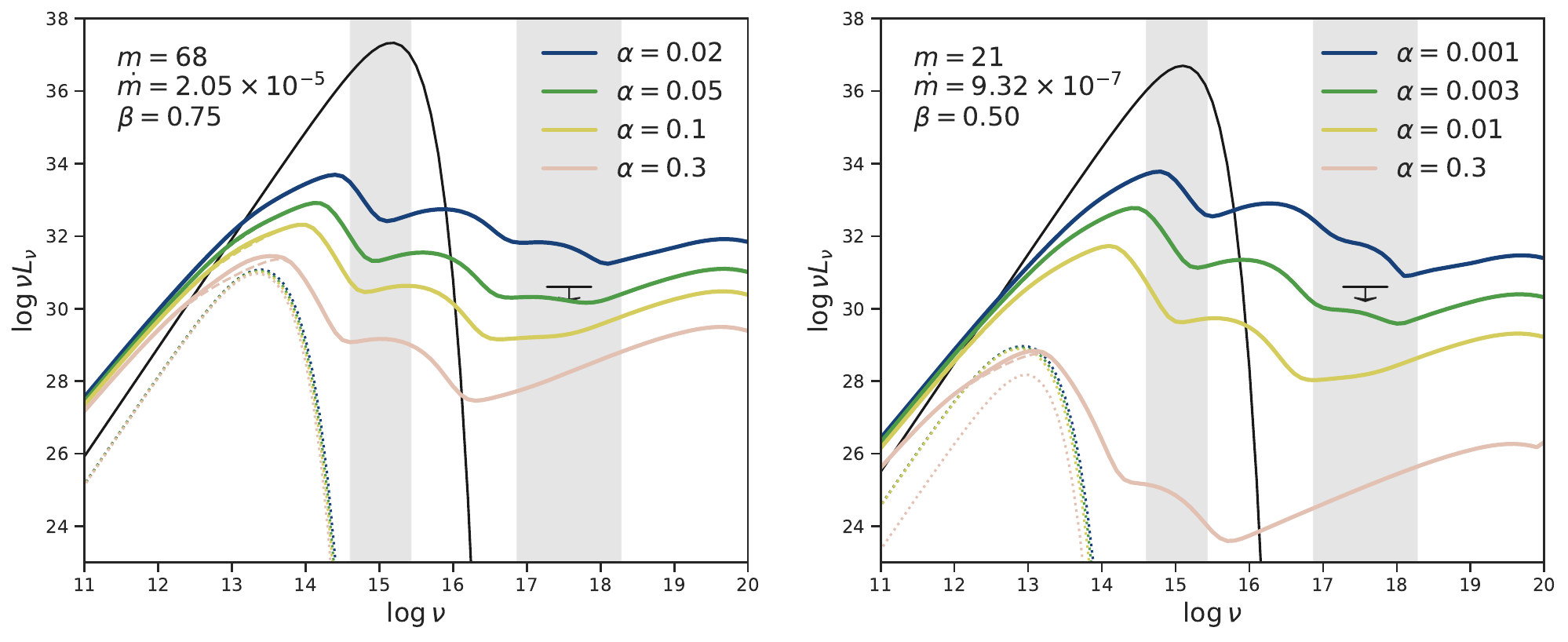}
\caption{
\textit{Left panel}: Emergent spectra of the model parameter $[m,\ \dot{m},\ \beta] = [68,\ 2.05\times10^{-5},\ 0.75]$ for different value of $\alpha$.
From the bottom up, the pink, yellow, green and blue solid lines are the emergent spectra for $\alpha=0.3, 0.1, 0.05$ and $0.02$ respectively. 
The pink, yellow, green and blue dotted and dashed lines are the emission from the truncated accretion disk and the emission of the inner ADAF for $\alpha=0.3, 0.1, 0.05$ and $0.02$, respectively. 
The black solid line denotes the blackbody radiation from the B-type star, with the effective temperature taken to be $T_\mathrm{eff}= 18000 K$ and the radius to be $R_\mathrm{B}=9R_\odot$.
\textit{Right panel}: Emergent spectra of the model parameter $[m,\ \dot{m},\ \beta] = [21,\ 9.32\times10^{-7},\ 0.50]$ for different value of $\alpha$. 
From the bottom up, the pink, yellow, green and blue solid lines are the emergent spectra for $\alpha=0.3, 0.01, 0.003$ and $0.001$ respectively. 
The pink, yellow, green and blue dotted and dashed lines are the emission from the truncated accretion disk and the emission of the inner ADAF for $\alpha=0.3, 0.01, 0.003$ and $0.001$, respectively. 
The black solid line denotes the blackbody radiation from the B-type star, with the effective temperature taken to be $T_\mathrm{eff}=$ 15300 K and the radius to be $R_\mathrm{B}=6R_\odot$. 
\label{fig:alphas1}}
\end{figure*}

\begin{deluxetable*}{ccccc}
\tabletypesize{\scriptsize}
\tablewidth{0pt} 
\tablecaption{Luminosities taking $[m,\ \dot{m},\ \beta] = [68,\ 2.05\times10^{-5},\ 0.75]$ and $[m,\ \dot{m},\ \beta] = [21,\ 9.32\times10^{-7},\ 0.50]$ respectively for different $\alpha$. 
Column (1) is the adopted $\alpha$.
Column (2) and column (3) are the bolometric luminosity ${L_\mathrm{bol, BH}}$ and the X-ray luminosity ${L_\mathrm{X, BH}}$ of the accretion flow (inner ADAF + outer truncated accretion disk) for different $\alpha$; Column (4) is the UV-optical luminosity ratio between the B-type star and the accretion flow around the BH $L_{\rm UV-Opt, B}/L_{\rm UV-Opt, BH}$.
\label{tab:para3}}
\tablehead{
\colhead{ } & \colhead{$\alpha$} & \colhead{$\log{L_\mathrm{bol, BH}}\ (\mathrm{erg/s})$} & 
\colhead{$\log{L_\mathrm{X, BH}}\ (\mathrm{erg/s})$} & \colhead{$\log({L_\mathrm{UV-Opt, B}/L_\mathrm{UV-Opt, BH}})$} \\
 & (1) & (2) & (3) & (4)
}
\startdata 
$m=68$& 0.3 & 31.77 & 28.88 & 8.01 \\
$\dot{m}=2.05\times10^{-5}$& 0.1 & 32.63 & 29.93 & 6.57 \\
$\beta=0.75$& 0.05 & 33.23 & 30.78 & 5.63 \\
& 0.02 & 34.04 & 32.17 & 4.27 \\
\hline
$m=21$& 0.3 & 29.11 & 25.73 & 11.66 \\
$\dot{m}=9.32\times10^{-7}$& 0.01 & 32.00 & 28.80 & 6.48 \\
$\beta=0.50$& 0.003 & 33.07 & 30.38 & 4.48 \\
& 0.001 & 34.12 & 32.35 & 3.00 \\
\enddata
\end{deluxetable*}

\section{discussion} \label{sec: discuss}


{\subsection{On the viscosity parameter $\alpha$ and the magnetic parameter $\beta$} \label{subsec:alpha_beta}
Viscosity is one of the most important physical processes in the black hole accretion theory, which controls the angular momentum transport and the heating of the matter in the accretion disk. \citet[][]{Shakura1973} proposed the so-called $\alpha$ description for the viscosity, in which the value of $\alpha$ is expected to be in the range of 0 to 1. The $\alpha$-description for the accretion flow has been widely used in different kinds of black hole accretion model, such as slim disk for higher mass accretion rates \citep[][]{1981ApJ...246..314A, 1986IAUS..119..371A,slim1988, 2000PASJ...52..133W,2000PASJ...52..499M,2001ApJ...549L..77W, 2013LRR....16....1A}, ADAF for lower mass accretion rates \citep[][]{Narayan1994, 1995ApJ...444..231N, 1995ApJ...452..710N, 1995Natur.374..623N, 1997ApJ...489..791M, 1998ApJ...492..554N, 2013ApJ...764....2Q, 2014A&A...563A.119B, 2014MNRAS.438.2804N, Yuan2014, 2015A&A...574A.133K, 2016ApJ...819..150F}, as well as in various hydrodynamical simulations of the accretion flow around BHs \citep[e.g., ][]{2019ApJ...880...67J, 2017MNRAS.469.4258T, 2016ApJ...826...23T, 2016MNRAS.456.3929S, 2015MNRAS.447...49S}. 
The magnetic parameter $\beta$ describes the strength of the magnetic field, which can play a role in the structure, dynamics and the radiation of the accretion flow to some extent depending on the magnitude of $\beta$ \citep[e.g.][]{Merloni2001,Liu2002,Liu2003, 2001PASJ...53L...1M, 2006AdSpR..38.1409L, 2006ApJ...641..103H, 2006ApJ...652L.113M, 2011MNRAS.418L..79T, 2012MNRAS.423.3083M, 2014MNRAS.439..503S, 2014MNRAS.441.3177M, 2019ARA&A..57..467B}.

As for the study in the present paper, the value of $\beta$ is constrained by comparing the width
of the observed $\rm H_{\alpha}$ emission line and the predicted truncation radius of the accretion disk from the disk evaporation model.
The value of $\alpha$ is constrained by comparing the upper limit of the X-ray luminosity 
given by \textit{Chandra} X-ray observatory and the theoretical X-ray luminosity of the accretion flow with a structure of an inner ADAF + an outer truncated accretion disk. We discuss the value of $\beta$ and $\alpha$ respectively as follows.

In Section \ref{sec:truncation}, based on G1 data, the value of $\beta$ is constrained to be  $\beta \sim 0.75$. Such a value of $\beta$ is consistent with the basic micro-physics in the disk evaporation model, in which the magnetic field in the corona above the accretion disk is believed to be weak, i.e., sub-equipartition ($\beta \gtrsim 0.5$) \citep[][]{2002A&A...392L...5M,Qian2007}. 
Meanwhile, it has been proved that the disk corona model can very smoothly connect
the outer disk-corona and the inner ADAF, which means that the corona above the accretion disk near 
the truncation radius of the accretion disk shares very similar properties with that of the adjacent ADAF
\citep[][]{Liu1999}.
The essence of ADAF is a kind of radiatively inefficient accretion flow, in which the magnetic field cannot be too strong. Otherwise, if there is a stronger magnetic field, i.e., $\beta < 0.5$, it is very possible that magnetic re-connection can be the dominant mechanism for heating the electrons in the ADAF. This will make ADAF very possibly to be radiatively efficient, which is inconsistent with the essence of the ADAF to be a kind of radiatively inefficient accretion flow \citep[][for review]{Ichimaru1977,Rees1982,Narayan1994,1997ApJ...489..791M,Narayan2008,Hawley2011,Qiaoetal2013,Yuan2014}. Based on G2 data, the value of $\beta$ is constrained to be  $\beta \sim 0.5$
(equipartition of the magnetic field), which still can be marginally accepted in the framework of the ADAF model and the disk evaporation model for the essence of a weaker magnetic field.


In Section \ref{sec:sp}, the value of $\alpha$ is constrained to be $\gtrsim 0.05$ for taking $m=68$, $\dot m=2.05\times10^{-5}$, and $\beta=0.75$, and $\gtrsim 0.003$ for taking $m=21$, $\dot m=9.32\times10^{-7}$, and $\beta=0.5$ respectively, which are all roughly consistent with both the numerical simulations \citep{2001ApJ...548..348H, 2002ApJ...566..164H,2013MNRAS.428.2255P, 2018ApJ...854....6H} and observations of other methods \citep[][]{Qiao2009,Qiao2018,2007MNRAS.376.1740K, 2019NewA...70....7M, 2021MNRAS.500.3976B, 2021ApJ...921..147C, 2022MNRAS.512.5269L}. Since in black hole binary systems, the X-ray emission is believed to originate from the accretion flow, it is expected that the value of $\alpha$ can be constrained in a more narrow range in future X-ray observations of LB-1.

}

 \subsection{On other quiescent black holes}

A group of binaries containing quiescent black holes have been discovered using radial velocity and astrometric methods, including AS 386 \citep{2018ApJ...856..158K}, NGC 3201 \#12560 and \#21859 \citep{2018MNRAS.475L..15G,2019A&A...632A...3G},
VFTS 243 \citep{2022NatAs...6.1085S}, HD 130298 \citep{2022A&A...664A.159M}, Gaia BH1 and  BH2 \citep{2023MNRAS.518.1057E, 2023MNRAS.521.4323E} and BH3 \citep{2024arXiv240410486G}, MWC 656 \citep{2014Natur.505..378C}, and 2M05215658 \citep{2019Sci...366..637T}.
Like LB-1, the nature of MWC 656 and 2M05215658 are also on debate \citep{2023A&A...677L...9J,2020Sci...368.3282V}.
No X-ray emission was detected for all these sources. The X-ray luminosity upper limits of VFTS 243, Gaia BH1 and BH2, MWC 656 and 2M05215658 were estimated to be 1.45$\times10^{32}$, 3$\times$10$^{30}$, 8$\times$10$^{29}$, 10$^{32}$, 4$\times$10$^{31}$ erg/s respectively. And the wind accretion rates of VFTS 243, and Gaia BH1 and BH2 were estimated to be $\approx$2$\times$10$^{-11}$, 2$\times$10$^{-17}$, 3$\times$10$^{-14}$ $M_\odot$/yr respectively. 
Such low mass accretion rates would definitely fall into the region of ADAF solution. Similar to the LB-1, it is expected that future X-ray observations can be used to constrain the value of viscosity parameter $\alpha$ in these binary systems.

\section{conclusion} \label{sec: conclude}
In this paper, in the scenario of a B-type main sequence star plus a BH for LB-1, we study the geometry of the accretion flow and calculate the corresponding emergent spectra. Specifically, we first calculate the mass-loss rate of the B-type star and the capture rate by the BH, with which as the initial mass accretion rate, we further calculate the truncation radius of the accretion disk in the framework of disk evaporation model and the corresponding emergent spectra of the accretion flow with an inner ADAF + an outer truncated accretion disk structure. 
Two groups of data, i.e., G1 \citep[]{2019Natur.575..618L} and G2 \citep[][]{Lennon2021} are used to test the physics of the accretion flow.
We found that both G1 and G2 data can well explain the observed width of the $\rm H_{\rm \alpha}$ emission line of LB-1 by calculating the truncation radius of the accretion disk with the disk evaporation model with proper magnetic parameter $\beta$.
We further constrain the viscosity parameter $\alpha$ by comparing the theoretical X-ray luminosity and the upper limit of the X-ray luminosity given by the \textit{Chandra} X-ray observatory. Finally, we discuss the possibility that future X-ray observations for the quiescent BHs could be 
a good probe for constraining the viscosity of the accretion flow, which can improve our understanding of the microphysics of the accretion flow around a BH. 

\begin{acknowledgments}
This work was supported by the National Natural Science Foundation of China (grant No. 12173048, 12333004, 11988101, 12273057), National Key R\&D Program of China (No. 2023YFA1607903). T.S. acknowledges support from the K.C. Wong Education Foundation.
\end{acknowledgments}

\bibliography{sample631}{}
\bibliographystyle{aasjournal}



\end{document}